\documentclass[reprint,amssymb, amsmath, aps, superscriptaddress, showpacs, footinbib,  prb]{revtex4-1}
\usepackage{graphicx}
\usepackage{color}

\newcommand{\eff}{\text{eff}}
\newcommand{\AFM}{\text{AFM}}
\newcommand{\FM}{\text{FM}}

\newcommand{\sub}{\text{sub}}

\newcommand{\Cels}{$^\circ$C}

\begin{document}

\title{(CuCl)LaTa$_2$O$_7$ and quantum phase transition\\ in the (CuX)LaM$_2$O$_7$ family (X = Cl, Br; M = Nb, Ta)}
\author{Alexander A. Tsirlin}
\email{altsirlin@gmail.com}
\affiliation{Max Planck Institute for Chemical Physics of Solids, N\"{o}thnitzer
Str. 40, 01187 Dresden, Germany}

\author{Artem M. Abakumov}
\affiliation{EMAT, University of Antwerp, Groenenborgerlaan 171, B-2020 Antwerp, Belgium}

\author{Clemens Ritter}
\affiliation{Institut Laue-Langevin, BP 156, F-38042 Grenoble, France}

\author{Helge Rosner}
\email{Helge.Rosner@cpfs.mpg.de}
\affiliation{Max Planck Institute for Chemical Physics of Solids, N\"{o}thnitzer
Str. 40, 01187 Dresden, Germany}
%\date{\today}

\begin{abstract}
We apply neutron diffraction, high-resolution synchrotron x-ray diffraction, magnetization measurements, electronic structure calculations, and quantum Monte-Carlo simulations to unravel the structure and magnetism of (CuCl)LaTa$_2$O$_7$. Despite the pseudo-tetragonal crystallographic unit cell, this compound features an orthorhombic superstructure, similar to the Nb-containing (CuX)LaNb$_2$O$_7$ with X = Cl and Br. The spin lattice entails dimers formed by the antiferromagnetic fourth-neighbor coupling $J_4$, as well as a large number of nonequivalent interdimer couplings quantified by an effective exchange parameter $J_{\eff}$. In (CuCl)LaTa$_2$O$_7$, the interdimer couplings are sufficiently strong to induce the long-range magnetic order with the N\'eel temperature $T_N\simeq 7$~K and the ordered magnetic moment of 0.53~$\mu_B$, as measured with neutron diffraction. This magnetic behavior can be accounted for by $J_{\eff}/J_4\simeq 1.6$ and $J_4\simeq 16$~K. We further propose a general magnetic phase diagram for the (CuCl)LaNb$_2$O$_7$-type compounds, and explain the transition from the gapped spin-singlet (dimer) ground state in (CuCl)LaNb$_2$O$_7$ to the long-range antiferromagnetic order in (CuCl)LaTa$_2$O$_7$ and (CuBr)LaNb$_2$O$_7$ by an increase in the magnitude of the interdimer couplings $J_{\eff}/J_4$, with the (CuCl)LaM$_2$O$_7$ (M = Nb, Ta) compounds lying on different sides of the quantum critical point that separates the singlet and long-range-ordered magnetic states.
\end{abstract}

\maketitle

\section{Introduction}
\label{sec:intro}
Quantum phase transitions are one of the hot topics in present-day solid-state physics. Experimental studies in this field enable a test of existing theories on quantum criticality as well as a direct access to exotic phases emerging in the vicinity of quantum critical points (QCPs). This task, however, remains highly challenging, because model systems amenable to the experimental study should lie close to the QCP and allow for a continuous tuning across the QCP. Such tuning can be achieved by applying external pressure or performing chemical substitutions. While external pressure facilitates continuous evolution of the system, the pressure cell complicates experimental work and restricts the set of applicable experimental techniques. Although measurements on chemically substituted samples are experimentally more feasible, the downside of the chemical approach is the inevitable disorder that may be detrimental for the physical effect under consideration.

Magnetic systems have become one of the main playgrounds for experimental and theoretical studies of quantum critical phenomena.\cite{ruegg2003,ruegg2004,*ruegg2008,sebastian2006,*sebastian2005,stone2006,*stone2007,zapf2006,*paduan2009} In insulating magnets, the effect of chemical substitution strongly depends on the position of the replaced atom and its contribution to magnetic couplings. Replacement of a ligand (anion) will usually disrupt the spin lattice, because ligand orbitals are an integral part of superexchange pathways that are responsible for the coupling. While systems with disrupted spin lattices are quite interesting on their own,\cite{[{For example: }][{}]manaka2008,*hong2010,cong2011,wulf2011} effects of bond randomness are highly unfavorable for experimental studies of quantum phase transitions in perfect, non-disrupted spin lattices. Cation replacement is a better method for tuning the system, because the spin lattice remains nearly homogeneous, whereas individual exchange couplings are only slightly modified owing to the change in lattice parameters, electrostatic fields, and crystal-field splittings. However, the effect of the cation replacement is often too weak to modify the magnetic ground state and drive the system across the QCP.

The (CuCl)LaNb$_{2-x}$Ta$_x$O$_7$ solid solutions are one of the promising systems with a possible quantum phase transition induced by the cation replacement.\cite{kitada2009} The Nb compound ($x=0$) reveals a singlet ground state with a spin gap,\cite{kageyama2005a,*kageyama2005b} whereas the Ta compound ($x=1$) is long-range antiferromagnetically ordered below $T_N\simeq 7$~K.\cite{kageyama2005c} The compositions with fractional $x$ values are intermediate between these two different regimes and presumably form a combination of the gapped singlet and gapless long-range-ordered magnetic phases.\cite{kitada2009,uemura2009}

Crystal structures of the (CuX)LaM$_2$O$_7$ compounds (X = Cl and Br, M = Nb and Ta) feature magnetic [CuX] layers sandwiched between non-magnetic [LaM$_2$O$_7$] perovskite slabs.\cite{kodenkandath1999,kodenkandath2001} Previous studies of the Nb-based systems established the formation of an orthorhombic superstructure related to the Jahn-Teller distortion of Cu$^{+2}$, the ordering of X atoms, and the tilting distortion within the [LaNb$_2$O$_7$] slabs.\cite{tsirlin2010a,tassel2010,hernandez2011,cubr} The resulting spin lattice entails spin dimers with a complex and still controversial pattern of interdimer couplings in (CuCl)LaNb$_2$O$_7$,\cite{tassel2010,tsirlin2010b} as well as strong interdimer couplings that trigger the long-range antiferromagnetic (AFM) order in (CuBr)LaNb$_2$O$_7$.\cite{cubr} Surprisingly, little is known about the crystal structure of (CuCl)LaTa$_2$O$_7$, and only its disordered version has been reported.\cite{kodenkandath2001} 

Because details of the low-temperature crystal structure and the nexus of the Ta and Nb compounds are crucial to understand the magnetic behavior of the (CuCl)LaNb$_{2-x}$Ta$_x$O$_7$ solid solutions, we performed an extensive structural study of (CuCl)LaTa$_2$O$_7$. The crystallographic results were further used for a microscopic analysis that demonstrated a close similarity between the Nb and Ta systems. Combining the crystallographic and microscopic information, we elucidate the differences in the atomic positions and the ensuing variation of the exchange couplings in the (CuX)LaM$_2$O$_7$ family. We provide a consistent, generalized, and quantitative description of the respective compounds, and address unresolved issues, such as the nature of magnetism of (CuCl)LaNb$_2$O$_7$.

%------------------------------------------------------------------------------------------------------------------------------
\section{Methods}
\label{sec:methods}
Powder samples of (CuCl)LaTa$_2$O$_7$ were prepared by a two-step procedure according to Ref.~\onlinecite{kodenkandath2001}. First, the RbLaTa$_2$O$_7$ precursor was obtained by firing a mixture of La$_2$O$_3$, Ta$_2$O$_5$, and 40~\% excess of Rb$_2$CO$_3$ at 800~\Cels\ for 12~h and 1050~\Cels\ for 12~h. The fused sample was washed with water, dried at 70~\Cels, and mixed with twice the equimolar amount of anhydrous CuCl$_2$ under argon atmosphere. This mixture was pressed into pellets, sealed in an evacuated quartz tube, and heated at 400~\Cels\ for 36~h. The sample was again washed with water to remove the excess CuCl$_2$ and the RbCl formed during the reaction, and dried overnight at 70~\Cels. Sample purity of the starting materials, the RbLaTa$_2$O$_7$ precursor, and the final product was controlled by powder x-ray diffraction (XRD).

Room-temperature XRD patterns were collected with the laboratory Huber G670 Guinier camera (CuK$_{\alpha1}$ radiation, $2\theta=3-100^{\circ}$ angle range, ImagePlate detector). High-resolution XRD data were further obtained at the ID31 beamline of European Synchrotron Radiation facility using a constant wavelength of $\lambda\simeq 0.4$~\r A and eight scintillation detectors, each preceded by a Si (111) analyzer crystal, in the angular range $2\theta=1-40$~deg. The powder sample was contained in a thin-walled borosilicate glass capillary that was spun during the experiment. The sample temperature was controlled by a He-flow cryostat (at 4.2~K), a liquid-nitrogen cryostream ($80-310$~K), and a hot-air blower ($350-750$~K).

Neutron diffraction data were collected at the high-resolution diffractometer D2B ($\lambda\simeq 1.595$~\r A, $Q=0.9-7.6$~\r A$^{-1}$, $T=10$~K) and the high-flux diffractometer D20 ($\lambda\simeq 2.417$~\r A, $Q=0.35-4.9$~\r A$^{-1}$, $T=1.5$~K and 40~K), both installed at the Institute Laue-Langevin (ILL, Grenoble, France). A 6~g powder sample used for all neutron measurements was loaded into a cylindrical vanadium container and cooled down with the standard Orange He-flow cryostat. Crystal and magnetic structures of (CuCl)LaTa$_2$O$_7$ were refined with \texttt{JANA2006}\cite{jana2006} and \texttt{FullProf}\cite{fullprof} programs, respectively.

Thermogravimetric analysis (TGA) was performed with STA409 Netzsch thermal balance in the $300-1070$~K temperature range in Ar atmosphere. The powder sample of (CuCl)LaTa$_2$O$_7$ was placed into a corundum crucible. 

The magnetic susceptibility was measured with the Quantum Design MPMS in the $4-380$~K temperature range in the applied magnetic field of 0.5~T.

To investigate electronic structure of (CuCl)LaTa$_2$O$_7$ and evaluate individual exchange couplings, we performed scalar-relativistic band structure calculations in the framework of density functional theory (DFT) with the local density approximation (LDA)\cite{pw92} and generalized gradient approximation (GGA)\cite{pbe96} for the exchange-correlation potential. The full-potential \texttt{FPLO9.01-35} code was used.\cite{fplo} Reciprocal space was sampled with a $k$ mesh of 256~points for the 48-atom crystallographic unit cell and 32 or 48 points for doubled 96-atom supercells. The convergence with respect to the $k$ mesh was carefully checked. The mean-field DFT+$U$ correction for strong correlation effects was applied. Further details of the computational procedure are described in Sec.~\ref{sec:dft}.

The magnetic model derived from the DFT calculations was further studied by quantum Monte Carlo (QMC) simulations using \texttt{loop}\cite{loop} and \verb|dirloop_sse|\cite{dirloop} algorithms of the \texttt{ALPS} simulation package.\cite{alps} Simulations were performed for finite lattices with periodic boundary conditions. The shapes and sizes of these lattices were adjusted to achieve an appropriate finite-size scaling for the staggered magnetization ($m_s$) and the N\'eel temperature ($T_N$), or to ensure the lack of finite-size effects for thermodynamic properties. Details of the simulations procedure are described in Sections~\ref{sec:simul} and~\ref{sec:model}.

%------------------------------------------------------------------------------------------------------------------------------
\section{Crystal structure}
\label{sec:structure}
The crystal structures of (CuCl)LaNb$_2$O$_7$ and (CuBr)LaNb$_2$O$_7$ are pseudotetragonal, with a tiny difference between the $a$ and $b$ lattice parameters.\cite{tsirlin2010a,tassel2010,hernandez2011,cubr} Although in the Nb compounds the orthorhombic splitting could be observed in high-resolution synchrotron XRD experiments, we failed to detect any signatures of a similar splitting in (CuCl)LaTa$_2$O$_7$. The structure refinement reported below (Table~\ref{tab:str}) suggests that the difference between the $a$ and $b$ parameters is indeed diminutively small and masked by the sizable reflection broadening, which is rather similar for different $hkl$ indices and signifies a sizable amount of defects and/or low particle size.

\begin{figure}
\includegraphics{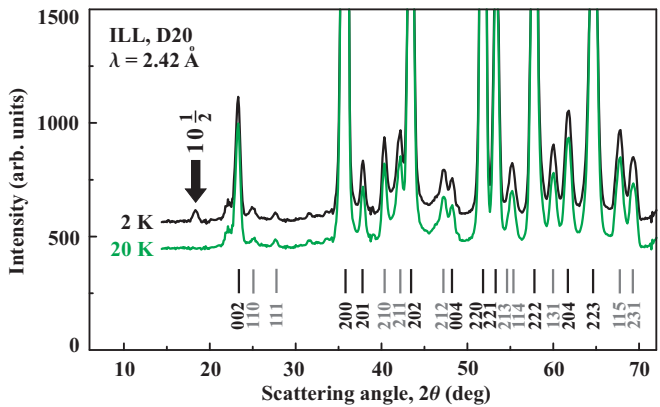}
\caption{\label{fig:neutron}
(Color online) Low-temperature neutron diffraction patterns of (CuCl)LaTa$_2$O$_7$ measured at 2~K and 20~K with the high-flux instrument D20. Black ticks indicate the reflections of the $a_{\sub}\times a_{\sub}\times c$ tetragonal unit cell reported in Ref.~\onlinecite{kodenkandath2001} (both $h$ and $k$ are even). Gray ticks show the superstructure reflections violating this unit cell (the $h$ and/or $k$ indices are odd).\cite{f1} The arrow denotes the magnetic reflection emerging below $T_N\simeq 7$~K.
}
\end{figure}
Owing to the large difference in the x-ray scattering from heavy Ta and light O atoms, neutron diffraction data were used for the structure refinement. Low-temperature neutron patterns (Fig.~\ref{fig:neutron}) revealed 10 superstructure reflections violating the simple tetragonal unit cell with $a_{\sub}\simeq 3.88$~\r A and $c_{\sub}\simeq 11.74$~\r A, as reported in Ref.~\onlinecite{kodenkandath2001}. Therefore, a four times larger $2a_{\sub}\times 2a_{\sub}\times c_{\sub}$ unit cell was used. In contrast to (CuBr)LaNb$_2$O$_7$,\cite{cubr} both superstructure reflections with even and odd $h+k$ were observed (Fig.~\ref{fig:neutron}), hence the $C$-centered unit cell reported for the Br compound could not be applied to (CuCl)LaTa$_2$O$_7$, and the primitive unit cell of (CuCl)LaNb$_2$O$_7$\cite{tsirlin2010a,tassel2010,hernandez2011} was used instead. Since the reflection conditions $h0l, h=2n$ and $0kl$, $k=2n$ are consistent with the $Pbam$ space group of (CuCl)LaNb$_2$O$_7$, the respective crystal structure was introduced in the refinement as the starting model.

\begin{figure}
\includegraphics{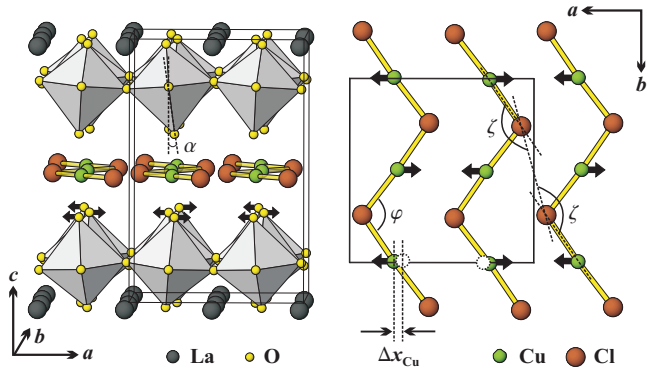}
\caption{\label{fig:str}
(Color online) Low-temperature crystal structure of (CuCl)LaTa$_2$O$_7$ with zigzag [CuCl] ribbons running along the $b$ direction. The arrows show the displacements of oxygen atoms upon the $a^0b^-c^0$ tilting distortion (left panel) and the resulting displacements of Cu atoms ($\Delta x_{\text{Cu}}$, right panel). The displacement $\Delta x_{\text{Cu}}$ is somewhat exaggerated for better visualization. The angle $\alpha$ measures the octahedral tilt, whereas $\varphi$ and $\zeta$ characterize the Cu--Cl--Cu ($J_1$) and Cu--Cl--Cl--Cu ($J_4$) superexchange pathways, respectively. Geometrical parameters are listed in Table~\ref{tab:dist}.
}
\end{figure}
Despite the overall successful refinement, the atomic displacement parameter (ADP) of Cl remained relatively high and possibly indicated the splitting of the Cl position, as previously observed for Cl atoms in single crystals of (CuCl)LaNb$_2$O$_7$ (Ref.~\onlinecite{hernandez2011}) and for Br atoms in (CuBr)LaNb$_2$O$_7$ (Ref.~\onlinecite{cubr}). However, the splitting of the Cl position neither reduced the ADP nor improved the refinement. No signatures of out-of-plane displacements of the Cl atoms were found either.

\begin{table}
\caption{\label{tab:str}
Atomic positions and isotropic atomic displacement parameters $U_{\text{iso}}$ (in $10^{-2}$~\r A$^2$) for (CuCl)LaTa$_2$O$_7$ according to the refinement of the neutron data at 1.8~K. Lattice parameters: $a=7.7663(5)$~\r A, $b=7.7640(3)$~\r A, $c=11.7374(5)$~\r A. Space group: $Pbam$. The $U_{\text{iso}}$ of oxygen atoms were refined as a single parameter. The error bars are based on the Rietveld refinement. 
}
\begin{ruledtabular}
\begin{tabular}{cccccc}
     & Position &  $x$      &   $y$     &   $z$     & $U_{\text{iso}}$ \\
 Cu  &   $4h$   & 0.7366(6) & 0.499(2)  & $\frac12$ &  0.47(7)         \\
 Cl  &   $4h$   & 0.5695(4) & 0.237(1)  & $\frac12$ &  1.4(1)          \\
 La  &   $4g$   & 0.0013(8) & 0.2566(7) &    0      &  0.08(5)         \\
 Ta  &   $8i$   & 0.7453(5) & 0.501(2)  & 0.8094(2) &  0.08(5)         \\
 O1  &   $4f$   & 0         & $\frac12$ & 0.8293(7) &  0.33(3)         \\
 O2  &   $8i$   & 0.2500(9) & 0.750(2)  & 0.8465(5) &  0.33(3)         \\
 O3  &   $8i$   & 0.2293(5) & $-0.001(2)$ & 0.6587(2) & 0.33(3)        \\
 O4  &   $4g$   & 0.7789(7) & 0.501(2)  & 0         &  0.33(3)         \\
 O5  &   $4e$   & $\frac12$ & $\frac12$ & 0.8592(7) &  0.33(3)         \\
\end{tabular}
\end{ruledtabular}
\end{table}
Refined atomic positions and main interatomic distances for (CuCl)LaTa$_2$O$_7$ are listed in Tables~\ref{tab:str} and~\ref{tab:dist}, respectively. Cu atoms have the typical four-fold coordination (CuO$_2$Cl$_2$ plaquette) with two short Cu--Cl bonds in the $ab$ plane and two Cu--O bonds along the $c$ direction. The short Cu--Cl bonds form zigzag ribbons along the $b$ direction, whereas the TaO$_6$ octahedra develop a tilting distortion according to the $a^0b^-c^0$ pattern.\footnote{Here, we refer to the conventional Glazer's notation, where $+/-$ stand for the in-phase/out-of-phase rotations about the respective crystallographic directions, and zero denotes the absence of the tilt.} In (CuCl)LaTa$_2$O$_7$, all [CuCl] ribbons have the same orientation, whereas in (CuBr)LaNb$_2$O$_7$ two types of ribbons related by a mirror symmetry are disordered in the averaged crystal structure (the space group $Cmmm$) and form alternating layers along the $c$ direction on the short-range scale (space group $Ibam$).\cite{cubr} A similar type of disorder may be responsible for the somewhat high ADP of Cl in (CuCl)LaTa$_2$O$_7$. However, the [CuCl] zigzag ribbons of different orientation could be present as single defects, only. These defects do not manifest themselves as the split Cl position in the averaged crystal structure probed by neutron diffraction.

\begin{table}
\caption{\label{tab:dist}
Selected geometrical parameters in the (CuX)LaM$_2$O$_7$ series: distances (in~\r A) and angles (in~deg). $\alpha$ denotes the tilting angle measured between the M--O3 bond and the $c$ direction. $\varphi$ and $\zeta$ describe the Cu--X--Cu ($J_1$) and Cu--X--X--Cu ($J_4$) superexchange pathways, as shown in Fig.~\ref{fig:str}. $\Delta x_{\text{Cu}}$ (in~\r A) measures the respective displacement of Cu atoms along the $a$ axis with respect to $x=\frac14$ and $\frac34$ (see Fig.~\ref{fig:str}). Crystallographic data for (CuCl)LaNb$_2$O$_7$ and (CuBr)LaNb$_2$O$_7$ are taken from Refs.~\onlinecite{hernandez2011} (neutron refinement) and~\onlinecite{cubr}, respectively.
}
\begin{ruledtabular}
\begin{tabular}{cccc}
         & (CuCl)LaTa$_2$O$_7$ & (CuCl)LaNb$_2$O$_7$ & (CuBr)LaNb$_2$O$_7$ \\
  Cu--O3 &    1.863(3)         &      1.863(6)       &      1.866(1)       \\
  Cu--X  &    2.38(2)          &      2.38(4)        &      2.49(3)        \\
  Cu--X  &    2.42(2)          &      2.39(4)        &      2.55(3)        \\
  M--O1  &    1.992(4)         &      1.969(7)       &      1.981(2)       \\
  M--O2  &    1.99(2)          &      2.00(4)        &      1.994(1)       \\
  M--O2  &    1.99(2)          &      1.97(4)        &      1.994(1)       \\
  M--O3  &    1.774(4)         &      1.767(5)       &      1.766(1)       \\
  M--O4  &    2.252(2)         &      2.252(4)       &      2.239(1)       \\
  M--O5  &    1.992(5)         &      2.007(8)       &      2.015(2)       \\\hline
  $\varphi$ & 108.3(4)         &      109.0(2)       &      101.8(2)       \\
  $\zeta$   & 163.8(4)         &      164.9(2)       &      158.4(2)       \\
  $\alpha$  &   4.0            &        5.7          &        4.8          \\
  $\Delta x_{\text{Cu}}$ & 0.104(4) &  0.126(2)      &      0.159(2)       \\
\end{tabular}
\end{ruledtabular}
\end{table}
It is further instructive to compare details of the atomic arrangement in the (CuX)LaM$_2$O$_7$ series. Table~\ref{tab:dist} shows that the [LaM$_2$O$_7$] perovskite slabs feature similar structures in all three compounds, with only a slight change in the tilting angle $\alpha$. Main changes are observed in the [CuX] layers, where the larger size of Br causes longer Cu--X distances and, consequently, the smaller Cu--X--Cu angle $\varphi$ compared to the Cl compounds. There is also a slight, yet significant difference between the Cu--X--Cu angles in (CuCl)LaNb$_2$O$_7$ and (CuCl)LaTa$_2$O$_7$.\footnote{Note that the refinement of the single-crystal XRD data for (CuCl)LaNb$_2$O$_7$ reveals an even larger Cu--Cl--Cu angle of 109.6(2)~deg (Ref.~\onlinecite{hernandez2011})} This difference is confirmed by DFT-based structure relaxations\footnote{Here, we use the GGA+$U$ method with the on-site Coulomb repulsion parameter $U_d=4.5$~eV. This choice has been justified in Refs.~\onlinecite{tsirlin2010a} and~\onlinecite{cubr}, where the results of the LDA-based and GGA-based relaxations are compared to the experiment.} that arrive at the Cu--Cl--Cu angles of 109.6~deg and 108.4~deg in the Nb and Ta compounds, respectively.

\begin{figure}
\includegraphics{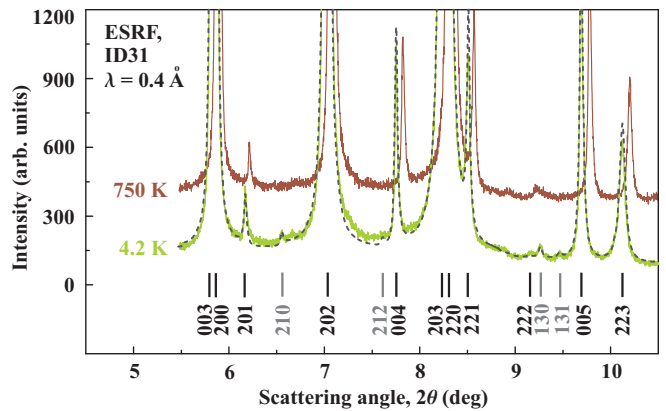}
\caption{\label{fig:xrd}
(Color online) High-resolution XRD patterns of (CuCl)LaTa$_2$O$_7$ measured at 4.2~K and 750~K. The dashed line shows the Rietveld refinement of the 4.2~K data. Black ticks indicate the reflections of the $a_{\sub}\times a_{\sub}\times c$ tetragonal unit cell reported in Ref.~\onlinecite{kodenkandath2001} (both $h$ and $k$ are even). Gray ticks show the superstructure reflections violating this unit cell (the $h$ and/or $k$ indices are odd). Note that the supercell reflections are few and weak, and disappear at 750~K. The tiny peak remaining at $2\theta\simeq 9.2$~deg at 750~K is the subcell reflection 222.
}
\end{figure}
The slight variation in the Cu--Cl--Cu angle can be explained by the different displacements of Cu atoms along the $a$ axis (see Table~\ref{tab:dist}). Previously, we have argued\cite{tsirlin2010a,cubr} that the Cu displacements are related to the tilting distortion in the perovskite slabs, because the O3 atoms shift along the $a$ direction upon the $a^0b^-c^0$ tilt and induce similar displacements of the Cu atoms, thus keeping the Cu--O3 bonds perpendicular to the Cu--Cl bonds in the $ab$ plane (see Fig.~\ref{fig:str}). In (CuCl)LaTa$_2$O$_7$, the smaller tilt leads to the smaller displacement $\Delta x_{\text{Cu}}$ and, therefore, results in the slightly reduced Cu--Cl--Cu angle. This mechanism elucidates the influence of non-magnetic perovskite slabs onto the magnetic [CuX] layers, and enables a deliberate modification of the magnetic layer via cation substitutions.

To investigate temperature evolution of the (CuCl)LaTa$_2$O$_7$ structure, we performed high-resolution XRD experiments in the $4.2-750$~K temperature range. In contrast to (CuCl)LaNb$_2$O$_7$ and (CuBr)LaNb$_2$O$_7$, the observation of structural changes was impeded by the fact that the superstructure reflections are barely visible in the XRD data (Fig.~\ref{fig:xrd}). Indeed, the superstructure originates from the displacements of light Cl and O atoms, while heavy Ta atoms do not contribute to the respective reflections. Additionally, the weak orthorhombic splitting could not be resolved because of the reflection broadening. Since the independent evaluation of the $a$ and $b$ parameters led to a very unstable refinement, we constrained $a=b$ in the whole temperature range. 

\begin{figure}
\includegraphics{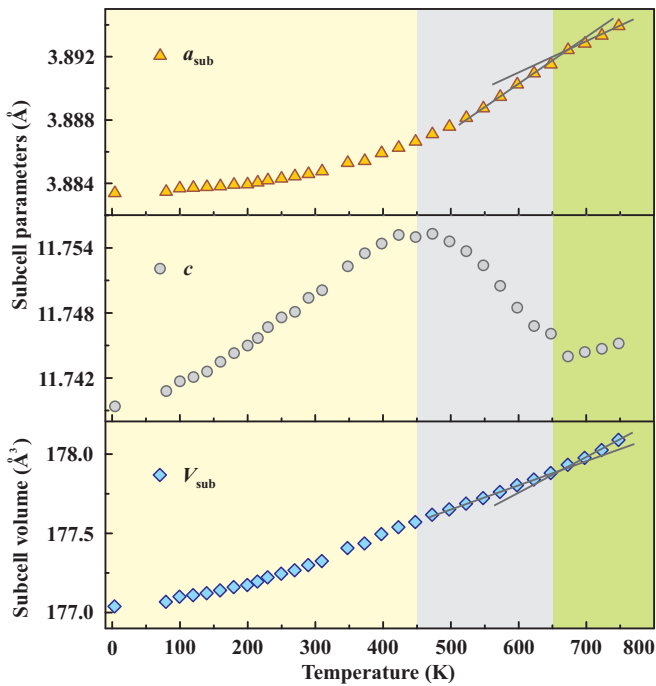}
\caption{\label{fig:cell}
(Color online) Temperature evolution of subcell lattice parameters of (CuCl)LaTa$_2$O$_7$. Different shadings denote the temperature ranges with dissimilar trends of the $c$ parameter. The transition at $T_2\simeq 650$~K is also evidenced by the slight change in the slope of the $a$ lattice parameter and of the subcell volume. Lines are guide-for-the-eye.
}
\end{figure}
A comparison of the x-ray patterns measured at 4.2~K and 750~K demonstrates that the superstructure reflections disappear upon heating. While the $a$ lattice parameter shows conventional thermal expansion with only slight changes in the slope, the peculiar temperature evolution of the $c$ parameter signifies structural phase transitions in (CuCl)LaTa$_2$O$_7$ (Fig.~\ref{fig:cell}). The increase in the $c$ value below 450~K is followed by a sharp decrease between 450~K and 650~K, and an eventual increase above 650~K. These changes are likely unrelated to chemical transformations, because the sample weight is nearly unchanged up to $800-850$~K\cite{supplement} and the heating/cooling processes are fully reversible, thus ruling out the possibility of a decomposition. Unfortunately, the x-ray data are insufficient for a precise structure determination owing to the very low intensity of the superstructure reflections. High-temperature neutron experiments would be required to establish the structural changes associated with the transitions. Nevertheless, the apparent analogy to (CuCl)LaNb$_2$O$_7$ (see Fig.~4 in Ref.~\onlinecite{tsirlin2010a}) suggests that the tilting distortion is eliminated upon the first structural transformation at $T_1\simeq 450$~K, while further heating destroys the ordered arrangement of Cl atoms above $T_2\simeq 650$~K. These transition temperatures are similar to $T_1\simeq 500$~K and $T_2=620-640$~K in (CuX)LaNb$_2$O$_7$.\cite{tsirlin2010a,cubr}

%------------------------------------------------------------------------------------------------------------------------------
\section{Magnetism}
\label{sec:magnetism}

\subsection{Microscopic model}
\label{sec:dft}
The close similarity between the crystal structures of (CuCl)LaNb$_2$O$_7$ and (CuCl)LaTa$_2$O$_7$ suggests that both compounds feature the same spin lattice with only a slight difference in the relevant microscopic parameters. To elucidate these subtle variations, we use DFT calculations that evaluate individual exchange couplings. 

\begin{figure}
\includegraphics{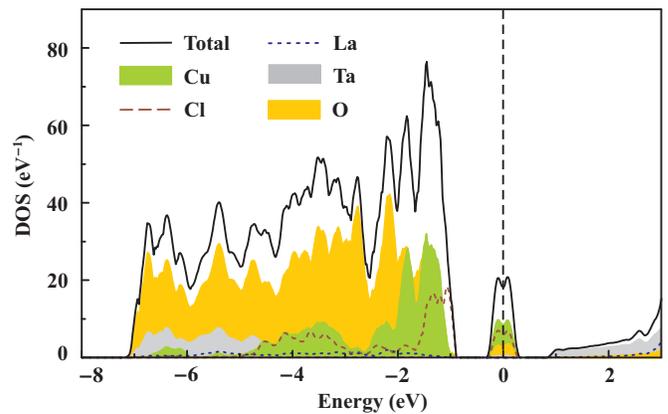}
\caption{\label{fig:dos}
(Color online) LDA density of states for (CuCl)LaTa$_2$O$_7$. The Fermi level is at zero energy.
}
\end{figure}

\begin{figure}
\includegraphics{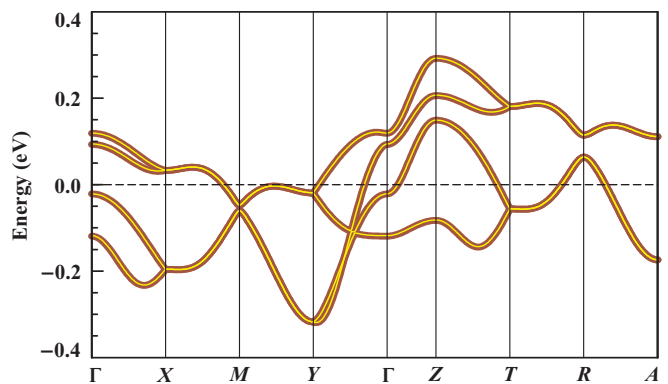}
\caption{\label{fig:band}
(Color online) LDA band structure (thin light lines) and the fit with the tight-binding model (thick dark lines). The Fermi level is at zero energy.
}
\end{figure}

The LDA band structure of (CuCl)LaTa$_2$O$_7$ is shown in Figs.~\ref{fig:dos} and~\ref{fig:band}. The valence band comprises oxygen $2p$ states between $-7$~eV and $-1$~eV as well as Cu $3d$ and Cl $3p$ states above $-5$~eV. The Fermi level is crossed by several bands forming a narrow complex between $-0.3$~eV and 0.3~eV. This band complex represents the $d_{x^2-y^2}$ crystal-field levels of Cu$^{+2}$ in the planar CuO$_2$Cl$_2$ environment, where local $x$ and $y$ axes are directed along the Cu--O and Cu--Cl bonds. Ta $5d$ bands are found above 0.85~eV, compared to $0.4-0.5$~eV for Nb $4d$ bands in (CuX)LaNb$_2$O$_7$. The spurious metallicity of the LDA energy spectrum should be ascribed to the strong underestimate of correlation effects in LDA. DFT+$U$ calculations result in realistic insulating spectra.

Similar to Refs.~\onlinecite{cubr} and~\onlinecite{tsirlin2010b}, we use two complementary approaches to the evaluation of exchange couplings. The first approach is based on the tight-binding fit of the LDA band structure, with the resulting hoppings $t_i$ introduced into an effective one-orbital Hubbard model featuring the on-site Coulomb repulsion $U_{\eff}$. Since $t_i\ll U_{\eff}$ and the bands are half-filled, low-lying (magnetic) excitations can be described by a Heisenberg model with the purely AFM exchange $J_i^{\AFM}=4t_i^2/U_{\eff}$. The $U_{\eff}=4$~eV value is used as a reasonable estimate for Cu$^{+2}$ (see Refs.~\onlinecite{cu2v2o7,cuncn,janson2012}).

The second approach involves local spin density approximation (LSDA)+$U$ calculations for a number of collinear spin configurations.\footnote{GGA+$U$ calculations lead to qualitatively similar results and feature the same ambiguity related to the AMF and FLL flavors of DFT+$U$.} Resulting energies are mapped onto a classical Heisenberg model to obtain full exchange couplings $J_i$. LSDA+$U$ introduces a mean-field correction for correlation effects in the Cu $3d$ shell, and provides accurate estimates of $J_i$ (see, e.g., Refs.~\onlinecite{janson2012,dioptase,pb2v3o9}) that are, however, rather sensitive to input parameters of the computational method. 

\begin{figure*}
\includegraphics{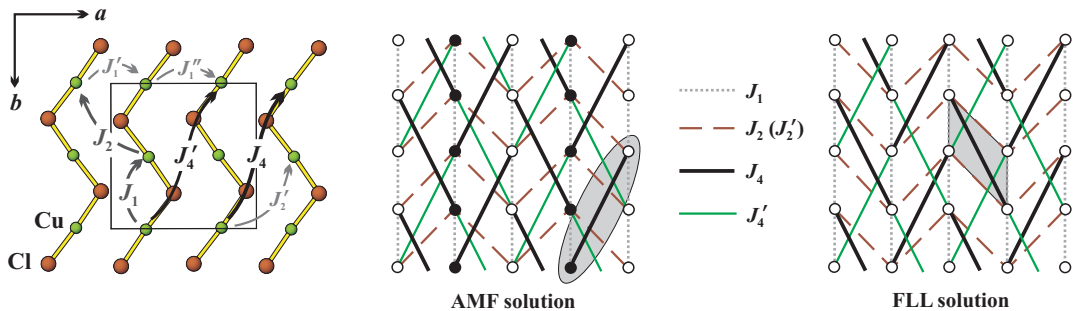}
\caption{\label{fig:lattice}
(Color online) Left panel: magnetic couplings in the [CuCl] layer. Middle and right panels: the respective spin lattices according to the AMF and FLL solutions in Table~\ref{tab:exchanges}. In the middle panel, the shading shows the spin dimer formed in (CuCl)LaNb$_2$O$_7$ by the coupling $J_4$, while empty and filled circles denote the stripe AFM order observed experimentally in (CuCl)LaTa$_2$O$_7$ and (CuBr)LaNb$_2$O$_7$. In the right panel, the shading indicates the Shastry-Sutherland plaquette with the intradimer coupling $J_4$ and frustrated interdimer couplings $J_1$ and $J_2'$, both FM.
}
\end{figure*}
Previously, we have argued\cite{cu2v2o7,volborthite,cuncn} that the on-site Coulomb repulsion parameter $U_d$ depends on the double-counting-correction (DCC) scheme that subtracts the part of the correlation energy already contained in LSDA. For each of the DCC schemes -- around-mean-field (AMF) and fully-localized-limit (FLL) -- the $U_d$ value has been adjusted with respect to the experimental Curie-Weiss temperature and saturation field of (CuCl)LaNb$_2$O$_7$, to obtain $U_d=4.5$~eV and 8.5~eV in AMF and FLL, respectively (see Ref.~\onlinecite{tsirlin2010b}). Here, we use the same $U_d$ values, because the local environment of Cu$^{+2}$ is essentially the same as in the Nb compound (see Table~\ref{tab:dist}). In (CuBr)LaNb$_2$O$_7$ (Ref.~\onlinecite{cubr}), the adjusted $U_d$ values are slightly different (5~eV and 12~eV, respectively), owing to the different nature of the ligands (Br instead of Cl) and different coordination environment. The on-site exchange parameter $J_d$ is fixed at 1~eV. An elaborate discussion on the role of the DCC in the evaluation of magnetic couplings within LSDA+$U$ can be found in Refs.~\onlinecite{cu2v2o7} and~\onlinecite{cdvo3}.

Table~\ref{tab:exchanges} summarizes the computed exchange couplings for all three compounds of the (CuX)LaM$_2$O$_7$ family. Here, we use $J_1$, $J_1'$, and $J_1''$ for inequivalent couplings between nearest neighbors, $J_2$ and $J_2'$ for couplings between next-nearest neighbors, etc, and $J_{\perp}$ for the interlayer coupling along the $c$ direction (Fig.~\ref{fig:lattice}). The hopping parameters $t_i$ are obtained from fits to the LDA band structure using Wannier functions adapted to specific orbital characters.\cite{wannier} These hopping parameters are directly related to AFM exchanges $J_i^{\AFM}$, whereas the $J_i$ values are an independent estimate of full exchange couplings that combine $J_i^{\AFM}$ with ferromagnetic (FM) contributions $J_i^{\FM}$. 

\begin{table*}
\caption{\label{tab:exchanges}
Calculated exchange couplings in the (CuX)LaM$_2$O$_7$ series. For each compound, the first column lists hopping parameters of the tight-binding model $t_i$ (in~eV), the second column contains the derived AFM contributions to the exchange $J_i^{\AFM}=4t_i^2/U_{\eff}$ (in~K) with $U_{\eff}=4$~eV, while the third and fourth columns contain full exchange couplings $J_i$ (in~K) obtained from LSDA+$U$ calculations with the AMF and FLL double-counting correction schemes. See text for details. The results for (CuBr)LaNb$_2$O$_7$ are taken from Ref.~\onlinecite{cubr}. The results for (CuCl)LaNb$_2$O$_7$ are calculated for the structural data from Ref.~\onlinecite{hernandez2011} and slightly differ from those reported in Ref.~\onlinecite{tsirlin2010b} previously.
}
\begin{ruledtabular}
\begin{tabular}{r@{\hspace{3em}}rrrr@{\hspace{3em}}rrrr@{\hspace{3em}}rrrr}
 & \multicolumn{4}{c}{(CuCl)LaTa$_2$O$_7$} & \multicolumn{4}{c}{(CuCl)LaNb$_2$O$_7$} & \multicolumn{4}{c}{(CuBr)LaNb$_2$O$_7$} \\
 & $t_i$ & $J_i^{\AFM}$ & $J_i$ & $J_i$ & $t_i$ & $J_i^{\AFM}$ & $J_i$ & $J_i$ & $t_i$ & $J_i^{\AFM}$ & $J_i$ & $J_i$ \\
             &          &    & AMF   &  FLL  &          &    &  AMF &  FLL  &          &      &  AMF  &  FLL  \\\hline
 $J_1$       & $-0.038$ & 17 & $-20$ & $-49$ & $-0.044$ & 23 & $-5$ & $-39$ &  0       &  0   & $-75$ & $-47$ \\
 $J_1'$      & 0.022    &  6 &   5   &  $-1$ &   0.018  & 4  &  4   &   1   & 0.045    &  24  &   31  &   7   \\
 $J_1''$     & 0.025    &  7 &   2   &  $-3$ &   0.025  & 7  & $-8$ & $-5$  & 0.034    &  13  &    1  &   4   \\
 $J_2$       & $-0.010$ &  1 &  26   &  $-3$ & $-0.008$ & 1  &  27  & $-3$  & $-0.047$ &  26  &   58  &  12   \\
 $J_2'$      & $-0.035$ & 14 &   9   & $-14$ & $-0.034$ & 13 &   2  & $-12$ & 0.011    &   1  & $-4$  & $-14$ \\
 $J_4$       & $-0.061$ & 43 &  69   &  40   & $-0.062$ & 45 &  67  &  45   & $-0.097$ & 110  &  144  &  54   \\
 $J_4'$      & $-0.042$ & 21 &  19   &  13   & $-0.042$ & 21 &  18  &  10   & $-0.036$ &  15  &   37  &  17   \\
 $J_{\perp}$ & $-0.035$ & 14 &  15   &  11   & $-0.038$ & 17 &  19  &  13   & $-0.028$ &   9  &   17  &   6   \\
\end{tabular}
\end{ruledtabular}
\end{table*}
Basic features of the spin lattice of the (CuX)LaM$_2$O$_7$ compounds have been extensively discussed in Refs.~\onlinecite{cubr} and~\onlinecite{tsirlin2010b}. Here, we only notice that the microscopic scenario is rather counter-intuitive because of the strongest AFM coupling between fourth neighbors. This unexpected result was originally derived from the inelastic neutron scattering experiment on (CuCl)LaNb$_2$O$_7$ (Ref.~\onlinecite{kageyama2005a,*kageyama2005b}) and later explained by the orthorhombic structural model with the efficient, albeit long, Cu--Cl--Cl--Cu superexchange pathway.\cite{tsirlin2009,tsirlin2010b,tassel2010} Another salient feature of the (CuX)LaM$_2$O$_7$ family is the sizable and nearly constant interlayer coupling $J_{\perp}$ mediated by the low-lying Nb $4d$ and Ta $5d$ states. An experimental signature of this effect is the large hyperfine coupling for Nb atoms, as probed with nuclear magnetic resonance (NMR).\cite{yoshida2007}

Despite certain discrepancies between the AMF and FLL results, exchange couplings evaluated by the same method enable a direct comparison to the structural data summarized in Table~\ref{tab:dist}. The variation in the Cu--X--Cu angle $\varphi$ has strong effect on the FM nearest-neighbor exchange $J_1$. The reduction in the angle from 109.0~deg in (CuCl)LaNb$_2$O$_7$ to 108.2~deg in (CuCl)LaTa$_2$O$_7$ and eventually to 101.8~deg in (CuBr)LaNb$_2$O$_7$ leads to a systematic increase in the absolute value of $J_1$, in agreement with Goodenough-Kanamori rules. Note that both $J_i^{\AFM}$ and $J_i^{\FM}=J_i-J_i^{\AFM}$ are changing with the angle. While $J_i^{\AFM}$ depends on the orbital overlap, $J_i^{\FM}$ is controlled by the Hund's coupling on the ligand site.\cite{mazurenko2007} 

The leading coupling $J_4$ remains nearly unchanged upon the Nb/Ta substitution, in agreement with similar $\zeta$ angles pertaining to the curvature of the Cu--Cl--Cl--Cu superexchange pathway (Table~\ref{tab:dist} and Fig.~\ref{fig:str}). Although in (CuBr)LaNb$_2$O$_7$ the $\zeta$ angle is notably decreased and the pathway becomes more curved, the intradimer exchange $J_4$ increases owing to the larger spatial extent of Br $4p$ orbitals compared to Cl $3p$. The AFM couplings $J_1',J_1''$, and $J_2$ are also enhanced.

The AMF and FLL calculations arrive at similar microscopic models that entail spin dimers formed by the leading coupling $J_4$ (Fig.~\ref{fig:lattice}, middle panel). The interdimer couplings include the FM nearest-neighbor interaction $J_1$, the AFM fourth-neighbor interaction $J_4'$, and the AFM interlayer interaction $J_{\perp}$. Next-nearest-neighbor couplings are also present, but their nature remains controversial. While AMF suggests a sizable AFM $J_2$ and a weakly FM $J_2'$, FLL puts forward the sizable FM $J_2'$ and a smaller $J_2$ that may be either FM or AFM depending on the compound. The AMF- and FLL-based scenarios are, therefore, different with respect to the possible frustration of the spin lattice. The AFM $J_2$ and negligible $J_2'$ in the AMF-based model are compatible with other couplings that altogether establish the stripe ordering pattern (parallel spins along $b$, antiparallel spins along $a$ and $c$, see the middle panel of Fig.~\ref{fig:lattice} and the inset to Fig.~\ref{fig:magstr}). The sizable FM $J_2'$ in the FLL-based model would, in contrast, frustrate this magnetic order and make the situation more complex.

Both AMF- and FLL-based models of (CuX)LaM$_2$O$_7$ have been considered in the literature. The AMF-based model (Fig.~\ref{fig:lattice}, middle panel) entails spin dimers with an intricate combination of non-frustrated interdimer couplings. Both thermodynamic and ground-state properties of this model can be precisely evaluated by QMC, as shown for (CuCl)LaNb$_2$O$_7$ (Ref.~\onlinecite{tsirlin2010b}), (CuBr)LaNb$_2$O$_7$ (Ref.~\onlinecite{cubr}), and (CuCl)LaTa$_2$O$_7$ alike (Sec.~\ref{sec:simul}). 

The FLL-based model may be thought of as a modified Shastry-Sutherland spin lattice, where spin dimers are coupled by frustrated FM couplings $J_1$ and $J_2'$ forming triangles (Fig.~\ref{fig:lattice}, right panel). This interpretation was proposed by Tassel~\textit{et al.},\cite{tassel2010} who performed FLL calculations only and basically ignored the sizable AFM couplings $J_4'$ and $J_{\perp}$ that are both missing in the Shastry-Sutherland geometry. Despite a subsequent theoretical work,\cite{furukawa2011} the application of the Shastry-Sutherland model to (CuCl)LaNb$_2$O$_7$ remains an obscure issue, because both thermodynamic and ground-state properties of the model, appropriately augmented with $J_4'$ and $J_{\perp}$, are hard to evaluate with a good precision. The QMC techniques fail, owing to the sign problem in a frustrated system, whereas exact diagonalization can not access sufficiently large finite lattices. Therefore, the application of the Shastry-Sutherland model to (CuCl)LaNb$_2$O$_7$ is, at least presently, not amenable to the experimental verification. 

In the following, we develop the microscopic description along the lines of Refs.~\onlinecite{cubr} and~\onlinecite{tsirlin2010b}, by focusing on the non-frustrated AMF-based model and exploring its properties with QMC. We further comment on the applicability of the Shastry-Sutherland model and its possible experimental verification (see Sec.~\ref{sec:model}).

%------------------------------------------------------------------------------------------------------------------------------
\subsection{Magnetic structure}
\label{sec:neutron}
Kitada~\textit{et al.}\cite{kitada2009} conjectured the stripe AFM ordering in (CuCl)LaTa$_2$O$_7$ based on the experimental observation of the strongest magnetic reflection, which is labeled as $10\frac12$ in Fig.~\ref{fig:neutron}. Here, we reconsider the magnetic structure of this compound in a powder neutron experiment that covers a wide angle range, thus giving access to all observable magnetic reflections and improving the estimate of the ordered magnetic moment. 

\begin{figure}
\includegraphics{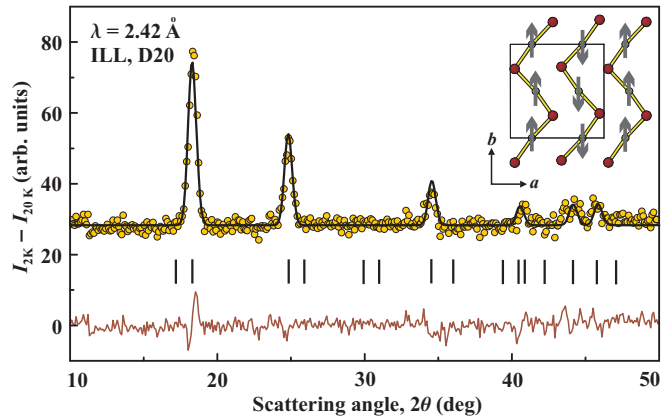}
\caption{\label{fig:magstr}
(Color online) Refinement of the magnetic structure. Experimental, calculated, and difference patterns are shown. Ticks denote the positions of magnetic reflections. The experimental and calculated patterns are offset from zero to ensure positive intensities. The inset shows the refined stripe AFM structure, with spins pointing along the $b$ direction.
}
\end{figure}
The neutron diffraction pattern measured at 2~K reveals one magnetic reflection only. The subtraction of the 20~K pattern, where only nuclear scattering is present, provides a full picture of magnetic scattering, with few additional weaker reflections at higher angles (Fig.~\ref{fig:magstr}). This subtracted pattern is refined as a purely magnetic phase. All magnetic reflections could be indexed with the propagation vector $\mathbf k=(0,0,\frac12)$, which together with the $4h$ position of Cu allows for several irreducible representations. However, only one of these representations allowed the successful refinement. 

The resulting magnetic structure features stripes of parallel spins along either $a$ or $b$ direction, with antiparallel spin arrangement along the $c$ direction and along the direction perpendicular to the stripes. Spins are directed along the stripes. Owing to the weak orthorhombic splitting, the neutron data are not sufficient to decide whether the stripes run along $a$ or $b$. However, the microscopic magnetic model gives clear indications for the stripes arranged along the $b$ direction according to the FM coupling $J_1$ (Fig.~\ref{fig:lattice} and Table~\ref{tab:exchanges}). This would also match the magnetic structure of (CuBr)LaNb$_2$O$_7$, where the arrangement of stripes along the [CuBr] zigzag ribbons ($b$ direction) is confirmed in an NMR experiment.\cite{cubr,yoshida2008}

The refined magnetic moment $\mu$ amounts to 0.53(1)~$\mu_B$, which is 0.2~$\mu_B$ lower than in (CuBr)LaNb$_2$O$_7$ ($\mu=0.72(1)$~$\mu_B$, Ref.~\onlinecite{cubr}) and in the previous estimate by Kitada~\textit{et al.} ($\mu=0.7(1)$~$\mu_B$).\cite{kitada2009} Phenomenologically, the lower magnetic moment is well in line with the lower N\'eel temperature $T_N\simeq 7$~K compared to $T_N\simeq 32$~K in the Br compound (see also $T_N/J_4$ in Table~\ref{tab:param}). The lower magnetic moment and the reduced N\'eel temperature suggest that in (CuCl)LaTa$_2$O$_7$ quantum fluctuations are enhanced compared to its (CuBr)LaNb$_2$O$_7$ analog.

%------------------------------------------------------------------------------------------------------------------------------
\subsection{Model simulations}
\label{sec:simul}
To quantify the magnetic model of (CuCl)LaTa$_2$O$_7$, we use QMC simulations for the Heisenberg spin Hamiltonian and the spin lattice derived from the DFT calculations (Table~\ref{tab:exchanges}, AMF solution). Unfortunately, the large number of inequivalent couplings prevents us from an independent evaluation of each exchange integral. We rather consider two relevant parameters, the intradimer coupling $J_4$ and the effective interdimer coupling $J_{\eff}$. The latter combines the interactions $J_1$, $J_2$, $J_4'$, and $J_{\perp}$, which are all of similar magnitude (see Table~\ref{tab:exchanges}). Therefore, we assume $-J_1=J_2=J_4'=J_{\perp}=j$ and $J_{\eff}=2|J_1|+2J_2+J_4'+2J_{\perp}=7j$, where the interdimer couplings are summed up according to their number per magnetic site (coordination number).

\begin{figure}
\includegraphics{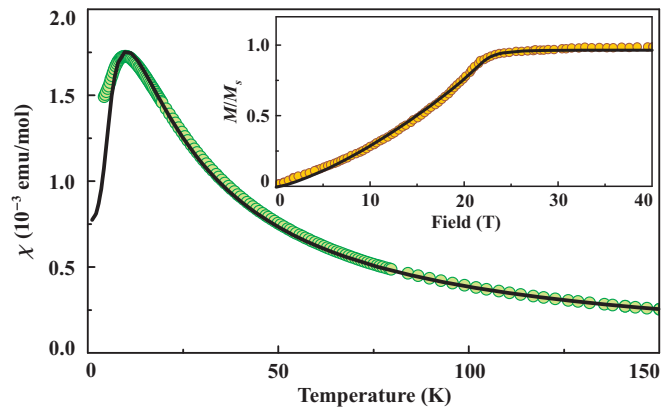}
\caption{\label{fig:simulations}
(Color online) Magnetic susceptibility of (CuCl)LaTa$_2$O$_7$ measured in the applied field of 0.5~T and the QMC fit with $J_4\simeq 16$~K, $J_{\eff}/J_4=1.6$. Deviations at low temperatures are likely related to an impurity contribution. The inset shows the magnetization isotherm measured at 1.3~K (experimental data are taken from Ref.~\onlinecite{kitada2009}) and the respective QMC fit for the same parameter set.
}
\end{figure}
The magnetic susceptibility of (CuCl)LaTa$_2$O$_7$ shows a broad maximum at $T^{\max}\simeq 10.5$~K followed by the AFM ordering at $T_N\simeq 7$~K, which is hardly visible in the magnetization data (Fig.~\ref{fig:simulations}) yet clearly identified by neutron measurements.\cite{kitada2009} Since the susceptibility does not show exponential decrease at low temperatures, the spin-gap scenario established for (CuCl)LaNb$_2$O$_7$ (Ref.~\onlinecite{kageyama2005a,*kageyama2005b}) can be safely excluded. The low-temperature behavior of the Ta compound is characteristic of a long-range ordered quantum antiferromagnet that develops the short-range magnetic order evidenced by the susceptibility maximum at $T^{\max}$, and eventually forms the long-range ordered state at $T_N<T^{\max}$.

The microscopic magnetic model of (CuCl)LaTa$_2$O$_7$ should allow for the long-range AFM order at low temperatures. While the single susceptibility curve can be reproduced with different model parameters, the simultaneous fit to the high-field magnetization isotherm taken from Ref.~\onlinecite{kitada2009} results in the unique solution with $J_4\simeq 16$~K and $J_{\eff}/J_4=1.6$ (Fig.~\ref{fig:simulations}).\footnote{Here, we used the $L\times L\times L$ finite lattice with $L=8$.} The fitted, powder-average $g$-value of $\bar g=2.15$ is in good agreement with the experimental estimates of $g_{\|}=2.13$ and $g_{\perp}=2.29$ reported for (CuCl)LaNb$_2$O$_7$.\cite{hernandez2011} 

\begin{table}
\caption{\label{tab:param}
Magnetic properties of the (CuX)LaM$_2$O$_7$ compounds: the N\'eel temperature ($T_N$), ordered magnetic moment $\mu$, saturation field ($\mu_0H_s$), intradimer coupling $J_4$, and the effective interdimer coupling $J_{\eff}$ (see text for details). The data for (CuCl)LaNb$_2$O$_7$ and (CuBr)LaNb$_2$O$_7$ are taken from Refs.~\onlinecite{tsirlin2010b} and~\onlinecite{cubr}, respectively.
}
\begin{ruledtabular}
\begin{tabular}{cccc}
                  & (CuCl)LaTa$_2$O$_7$ & (CuCl)LaNb$_2$O$_7$ & (CuBr)LaNb$_2$O$_7$ \\
 $T_N/J_4$        &         0.44        &         --          &        0.67         \\
 $\mu$ ($\mu_B$)  &         0.53        &         --          &        0.72         \\
 $\mu_0H_s$ (T)   &         22          &         30          &        84           \\
 $J_4$ (K)        &         16          &         25          &        48           \\ 
 $J_{\eff}/J_4$   &         1.6         &         0.4         &        3.0          \\
\end{tabular}
\end{ruledtabular}
\end{table}
Similar to (CuCl)LaNb$_2$O$_7$ and (CuBr)LaNb$_2$O$_7$, there is a certain ambiguity in the choice of interdimer couplings, because the overall coupling can be redistributed between weak AFM exchanges $J_2,J_4'$, and $J_{\perp}$. Therefore, only the effective interdimer coupling is determined with sufficient accuracy. The estimated $J_{\eff}/J_4$ ratios (Table~\ref{tab:param}) show clearly that the interdimer couplings are enhanced in (CuCl)LaTa$_2$O$_7$ compared to its Nb-containing counterpart, whereas (CuBr)LaNb$_2$O$_7$ features an even larger interdimer exchange. This evolution of the spin lattice underlies the transition from the gapped spin-singlet ground state in (CuCl)LaNb$_2$O$_7$ to the stripe AFM order in the other two compounds, as further discussed in Sec.~\ref{sec:model}. It also explains the increase in the N\'eel temperature and ordered magnetic moment in (CuBr)LaNb$_2$O$_7$ compared to (CuCl)LaTa$_2$O$_7$ (see Table~\ref{tab:param}). 

\begin{figure}
\includegraphics{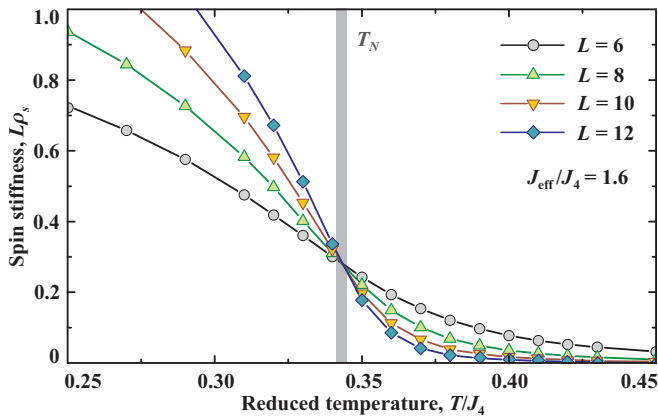}
\caption{\label{fig:stif-1}
(Color online) Temperature dependence of the spin stiffness $L\rho_s$ calculated for $L\times L\times L$ finite lattices of different size $L$ at $J_{\eff}/J_4=1.6$. The shaded bar identifies the N\'eel temperature $T_N/J_4\simeq 0.34$ corresponding to $T_N\simeq 5.5$~K for $J_4=16$~K.
}
\end{figure}
The change in the intradimer coupling $J_4$ is illustrated by the experimental values of the saturation field $\mu_0H_s$ (Table~\ref{tab:param}). Thus, the lower $H_s$ in (CuCl)LaTa$_2$O$_7$ compared to its Nb-containing counterpart stems from the lower $J_4$, even though the interdimer couplings are increased. The saturation field of (CuBr)LaNb$_2$O$_7$ is, by contrast, about three times higher because of the larger $J_4$ and the pronounced increase in $J_{\eff}$. Altogether, our estimates of exchange couplings -- $J_4$ and $J_{\eff}$ -- are well in line with the experimental observations.

To perform a more elaborate comparison between the proposed model and experiment, we estimated the ordered magnetic moment $\mu$ for the spin lattice with $J_{\eff}/J_4=1.6$. Following the procedure described in Ref.~\onlinecite{sandvik1997}, we computed static structure factors at the experimental propagation vector $\mathbf k=(0,0,\frac12)$ for $L\times L\times L$ finite lattices with $L\leq 12$, and performed the finite-size scaling. This way, the staggered magnetization $m_s=0.681$~$\mu_B$ is obtained. To compare this model result with the experimental $\mu$, one should additionally take into account the spin-orbit coupling quantified by the $g$-value ($\mu=gSm_s$ with $S=\frac12$) and the hybridization with the ligands. According to the LSDA+$U$ calculations, the hybridization reduces the magnetic moment by $25-30$~\%,\footnote{The magnetic moment of Cu is 0.68~$\mu_B$ in AMF at $U_d=4.5$~eV and 0.74~$\mu_B$ in FLL at $U_d=8.5$~eV.} thus leading to $\mu=0.51-0.55$~$\mu_B$ in remarkable agreement with the experimental estimate of 0.53~$\mu_B$.

The N\'eel temperature can be evaluated from the temperature dependence of the spin stiffness $\rho_s$. The spin stiffness is close to zero in the paramagnetic state (above $T_N$) and reaches a finite value in the long-range-ordered state (below $T_N$). In the vicinity of $T_N$, a steep increase in $\rho_s$ is observed. The scaling properties\cite{sengupta2003} of $\rho_s$ suggest that for a three-dimensional spin lattice the quantity $L\rho_s$ is independent of the lattice size $L$ at the transition temperature $T_N$. Therefore, $T_N$ can be determined as the crossing point of $L\rho_s(T)$ curves calculated for finite lattices with different $L$. The scaling procedure is shown in Fig.~\ref{fig:stif-1} and results in $T_N/J_4\simeq 0.34$ and $T_N\simeq 5.5$~K in good agreement with the experimental $T_N\simeq 7$~K.\cite{kitada2009}

%------------------------------------------------------------------------------------------------------------------------------
\section{General magnetic model and Discussion}
\label{sec:model}

\begin{figure}
\includegraphics{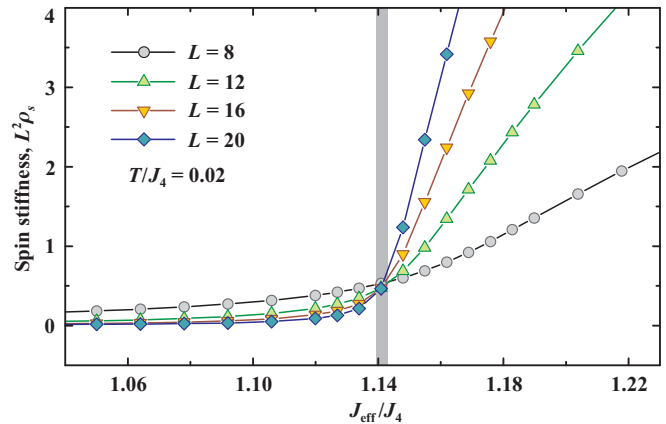}
\caption{\label{fig:stif-2}
(Color online) Scaling procedure for locating the QCP. The spin stiffness ($\rho_s$) multiplied by the squared lattice size ($L^2$) is plotted as a function of the effective interdimer coupling $J_{\eff}/J_4$ at the constant temperature of $T/J_4=0.02$. The crossing point indicates the QCP at $J_{\eff}/J_4\simeq 1.14$ separating the spin-singlet and long-range-ordered ground states. For the calculations, the $L\times L\times L/2$ finite lattices are used.
}
\end{figure}
The accurate microscopic description of (CuCl)LaNb$_2$O$_7$ (Ref.~\onlinecite{tsirlin2010b}), (CuBr)LaNb$_2$O$_7$ (Ref.~\onlinecite{cubr}), and (CuCl)LaTa$_2$O$_7$ (Sec.~\ref{sec:simul}) enables us to develop a common magnetic model for the (CuX)LaM$_2$O$_7$ family. Although the three compounds slightly differ in the coupling regime (e.g., $J_1'$ and $J_1''$ manifest themselves in the Br compound, while remaining weak in the Cl-based systems, see Table~\ref{tab:exchanges} and Ref.~\onlinecite{cubr}), their properties are well captured by the same spin lattice with the intradimer coupling $J_4$ and the assumption of equal couplings $-J_1=J_2=J_4'=J_{\perp}=j$ that contribute to the effective interdimer coupling $J_{\eff}=7j$. The magnetic behavior of this model is fully determined by the $J_{\eff}/J_4$ ratio. At low $J_{\eff}$, the spin dimers are weakly coupled and develop a spin gap in a spin-singlet (dimer) state without the long-range magnetic order. Large interdimer couplings close the spin gap and establish the stripe AFM order below $T_N$.

\begin{figure}
\includegraphics{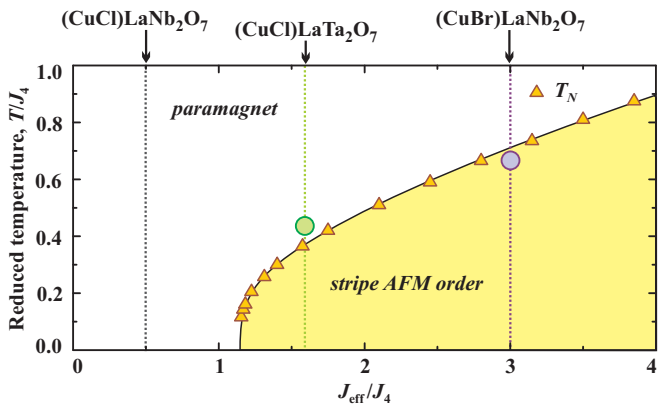}
\caption{\label{fig:diagram-1}
(Color online) Magnetic phase diagram of the (CuX)LaM$_2$O$_7$ compounds. Triangles show calculated N\'eel temperatures $T_N$ for different $J_{\eff}/J_4$ ratios and denote the boundary of the magnetically ordered phase. Dotted lines indicate the parameter regimes of three experimentally studied compounds, whereas large circles mark the experimental $T_N$ for (CuCl)LaTa$_2$O$_7$ and (CuBr)LaNb$_2$O$_7$, where the stripe AFM order is observed.
}
\end{figure}
The regions of the spin-singlet ground state and stripe AFM order are separated by a QCP. The precise position of the critical point can be determined from simulations of the spin stiffness, similar to the estimate of the N\'eel temperature given in Sec.~\ref{sec:simul}. The spin stiffness is close to zero in the spin singlet state, and reaches a finite value in the long-range-ordered state. Therefore, upon crossing the QCP the $\rho_s$ quantity changes abruptly, and for a three-dimensional spin system the crossing point of the $L^2\rho_s(T)$ curves calculated for different lattice size $L$ provides the position of the QCP at $J_{\eff}/J_4\simeq 1.14$ (see Fig.~\ref{fig:stif-2}).\cite{albuquerque2008}

At low $J_{\eff}/J_4$, the spin-singlet state continuously transforms into the high-temperature paramagnetic regime upon heating. By contrast, the elimination of the stripe AFM state is accompanied by an abrupt phase transition at $T_N$, which determines the phase boundary. The N\'eel temperature as a function of $J_{\eff}/J_4$ is evaluated by QMC simulations for different $J_{\eff}/J_4$ ratios to obtain the phase diagram presented in Fig.~\ref{fig:diagram-1}. The (CuX)LaM$_2$O$_7$ compounds are placed on this diagram according to their $J_{\eff}/J_4$ ratios determined from QMC fits to the magnetic susceptibility and high-field magnetization (Fig.~\ref{fig:simulations}). The comparison to the experimental $T_N$ confirms the accuracy of our model that properly captures the reduction in $T_N$ in (CuCl)LaTa$_2$O$_7$ compared to (CuBr)LaNb$_2$O$_7$. 

\begin{figure}
\includegraphics{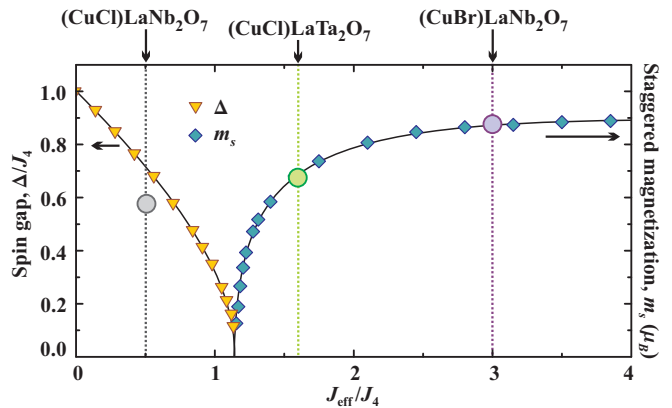}
\caption{\label{fig:diagram-2}
(Color online) Spin gap $\Delta$ (left axis) and staggered magnetization $m_s$ (right axis) in the microscopic magnetic model of the (CuX)LaNb$_2$O$_7$ compounds. Dotted lines indicate the parameter regimes of three experimentally studied compounds, whereas large circles are experimental results. See text for details.
}
\end{figure}
Similar to $T_N$, other parameters obtained from the experiment can be correlated with the $J_{\eff}/J_4$ ratio. In Fig.~\ref{fig:diagram-2}, we plot the spin gap for the spin-singlet state at low $J_{\eff}/J_4$ and the staggered magnetization for the stripe AFM state at high $J_{\eff}/J_4$. Upon increasing the interdimer couplings, the spin gap is systematically reduced compared to $\Delta=J_4$ for an isolated dimer at $J_{\eff}=0$, and eventually vanishes at the QCP. At higher $J_{\eff}/J_4$, the staggered magnetization appears. It is worth noting that the trends for $T_N$ and $m_s$ are somewhat dissimilar. While $T_N$ depends on the overall energy of the exchange couplings and linearly increases with the interdimer couplings above $J_{\eff}/J_4\simeq 1.5$, the staggered magnetization approaches the saturated value of $m_s\simeq 0.89$~$\mu_B$. This value is 11~\% lower than the maximum magnetic moment of 1~$\mu_B$ for spin-$\frac12$, and represents the effect of AFM spin fluctuations as well as residual quantum effects related to the low magnetic moment on the Cu site. 

Experimental spin gap and staggered magnetizations are in good agreement with our general model (see Fig.~\ref{fig:diagram-2}). However, this comparison is less straight-forward than in the case of $T_N$, because $m_s$ is usually different from the ordered magnetic moment $\mu$ measured by neutron diffraction. Therefore, the experimental values of $\mu$ should be scaled with the $g$-value and Cu--ligand hybridization, as explained in Sec.~\ref{sec:simul}. While the experimental $\mu$ value is not a direct measure of $J_{\eff}/J_4$, the changes in $\mu$ reflect the evolution of underlying exchange couplings and, e.g., manifest the reduction in $J_{\eff}/J_4$ in (CuCl)LaTa$_2$O$_7$ compared to (CuBr)LaNb$_2$O$_7$. The spin gap $\Delta$ can be estimated by different methods, with inelastic neutron scattering and high-field magnetization often leading to different results as, e.g., in (CuCl)LaNb$_2$O$_7$.\cite{kageyama2005a,*kageyama2005b} In our QMC simulations, we estimated $\Delta$ from the first critical field $H_{c1}$, where the magnetization of the system starts increasing. Therefore, for the experimental estimate we used $\Delta\simeq 14$~K obtained from $\mu_0H_{c1}\simeq 10.5$~T in (CuCl)LaNb$_2$O$_7$. 

Altogether, our general magnetic model elucidates the differences between the compounds of the (CuX)LaM$_2$O$_7$ family, and fully conforms to microscopic trends established by DFT calculations (Table~\ref{tab:exchanges}). The replacement of Nb by Ta modifies the tilting angle of the MO$_6$ octahedra within the [LaM$_2$O$_7$] slabs, thus slightly changing the Cu--Cl--Cu ($\varphi$) and Cu--Cl--Cl ($\zeta$) angles in the $ab$ plane, enhancing the FM coupling $J_1$, and reducing the AFM coupling $J_4$. The latter effect, although not well visible in the DFT results, is unequivocally established experimentally (Table~\ref{tab:param}). Although small at first glance, this alteration of the spin lattice is sufficient to change the magnetic ground state, because both (CuCl)LaNb$_2$O$_7$ and (CuCl)LaTa$_2$O$_7$ are relatively close to the QCP in terms of the $J_{\eff}/J_4$ ratio (see Figs.~\ref{fig:diagram-1} and~\ref{fig:diagram-2}). Since (CuBr)LaNb$_2$O$_7$ lies further away from the QCP, the Nb/Ta substitution should have little effect on the magnetism. Indeed, (CuBr)LaTa$_2$O$_7$ undergoes the AFM ordering at $T_N\simeq 35$~K, which is similar to $T_N\simeq 32$~K in (CuBr)LaNb$_2$O$_7$.\cite{kageyama2005c}

Our results do not support the earlier conjecture by Kitada~\textit{et al.}\cite{kitada2009} who ascribed the qualitative difference between (CuCl)LaNb$_2$O$_7$ and (CuCl)LaTa$_2$O$_7$ to a different regime of interlayer couplings mediated by the Nb and Ta atoms. We rather show that the interlayer coupling $J_{\perp}$ is essentially unchanged, whereas the different tilting angle of the NbO$_6$ and TaO$_6$ octahedra leads to slight changes in the [CuCl] layers, thus modifying the intralayer exchange couplings. In (CuX)LaM$_2$O$_7$, the variable magnetic behavior is related to structural changes in the $ab$ plane, although the spin lattices are basically three-dimensional, with a sizable interlayer coupling $J_{\perp}$ observed in all three compounds (Table~\ref{tab:exchanges}).

The (CuX)LaM$_2$O$_7$ compounds strongly resemble the family of spin-dimer ACuX$_3$ halides (A = K, Tl; X = Cl, Br), where quantum phase transitions between the gapped spin-singlet state and long-range AFM order were extensively studied experimentally. For example, TlCuCl$_3$ features a spin gap $\Delta/J\simeq 0.6$ at ambient pressure and transforms into the long-range-ordered antiferromagnet above $0.05-0.1$~GPa.\cite{ruegg2004,*ruegg2008,oosawa2003,*oosawa2004,*goto2004} However, in the ACuX$_3$ family the ordered ground state can not be stabilized at ambient pressure, unless an additional structural distortion is present, as experimentally found for NH$_4$CuCl$_3$ (Ref.~\onlinecite{ruegg2004b,*matsumoto2003}). An advantage of the (CuX)LaM$_2$O$_7$ family is the possibility to stabilize both spin-singlet and long-range-ordered phases at ambient pressure. On the downside, single crystals of (CuX)LaM$_2$O$_7$ are exceedingly difficult to prepare,\cite{hernandez2011} and the exact pattern of interdimer couplings has not been established experimentally. The spin lattice of ACuX$_3$ is similarly complex, yet amenable to the experimental study by inelastic neutron scattering on single crystals.

The lack of a direct experimental information on the interdimer couplings resulted in a controversy regarding the precise microscopic magnetic model of (CuCl)LaNb$_2$O$_7$ and the frustration of interdimer couplings in this compound (compare Refs.~\onlinecite{tassel2010} and~\onlinecite{tsirlin2010b}). Here, we elaborated on the non-frustrated version of the model (Fig.~\ref{fig:lattice}, middle panel),\cite{tsirlin2010b} and demonstrated its applicability to the whole (CuX)LaM$_2$O$_7$ family. This model not only captures the qualitative effect of different magnetic ground states depending on the interdimer couplings, but also provides a decent quantitative description, as shown in Figs.~\ref{fig:diagram-1} and~\ref{fig:diagram-2}. Considering the enigmatic nature of early experimental results on (CuCl)LaNb$_2$O$_7$ and related compounds,\cite{kageyama2005a,*kageyama2005b} as well as futile attempts to understand the underlying physics in terms of the square-lattice spin models,\cite{oba2006} our results are a remarkable accomplishment that confirms the excellent potential of DFT and precise numerical simulations in the microscopic evaluation of complex spin systems. Nevertheless, one has to keep in mind inevitable limitations of this approach that does not allow us to decide unambiguously between the models with the frustrated and non-frustrated interdimer couplings (compare the right and middle panels of Fig.~\ref{fig:lattice}).

In Sec.~\ref{sec:dft}, we have argued that the FM Shastry-Sutherland model proposed by Tassel~\textit{et al.}\cite{tassel2010} should be augmented by additional AFM interdimer couplings, and remains inaccessible to quantitative verification with feasible experimental techniques. Qualitatively, this model could be consistent with the quantum phase transition toward the stripe AFM ordering in (CuCl)LaTa$_2$O$_7$ and (CuBr)LaNb$_2$O$_7$, provided that one of the frustrating FM interdimer couplings becomes sufficiently weak or even AFM.\cite{furukawa2011} Therefore, the magnetic frustration is not a generic feature of the whole (CuX)LaM$_2$O$_7$ family and may only affect the behavior of (CuCl)LaNb$_2$O$_7$, although even this conjecture requires further experimental verification. Future inelastic neutron scattering experiments on single crystals should elucidate to what extent the frustration and the Shastry-Sutherland-type physics are relevant for (CuCl)LaNb$_2$O$_7$.

A natural reason for further experimental work on the (CuX)LaM$_2$O$_7$ family is the fact that the two Cl-containing compounds lie on different sides of the QCP. We have shown that a tiny structural effect related to the Nb/Ta replacement is responsible for the change in the magnetic ground state. A similar change could be induced by hydrostatic pressure or chemical substitution. The available experimental data on the (CuCl)LaNb$_{2-x}$Ta$_x$O$_7$ solid solutions\cite{kitada2009} suggest that the system separates into the long-range-ordered and spin-singlet phases. It would be interesting to understand whether or not this separation originates from a chemical inhomogeneity, and whether or not an improved synthetic procedure could facilitate the continuous evolution of the system and the experimental access to the proposed QCP.

In summary, we extensively characterized (CuCl)LaTa$_2$O$_7$ with respect to its crystal structure, magnetic behavior, and microscopic magnetic model. Our results place (CuCl)LaTa$_2$O$_7$ between (CuCl)LaNb$_2$O$_7$ and (CuBr)LaNb$_2$O$_7$, and evidence sizable quantum fluctuations in this compound. The enhanced quantum fluctuations indicate the evolution toward the quantum critical point that separates the spin-singlet (dimer) and long-range ordered ground states. The Nb/Ta replacement changes the tilting angles of the respective MO$_6$ octahedra and consequently alters the positions of Cu atoms in the [CuCl] layers, thus enhancing the nearest-neighbor ferromagnetic coupling $J_1$, and reducing the intradimer coupling $J_4$. This slight structural change is sufficient to induce a dramatic change in the magnetic ground state and drive the system across the quantum critical point. Therefore, both (CuCl)LaNb$_2$O$_7$ and (CuCl)LaTa$_2$O$_7$ as well as their solid solutions may be interesting materials for experimental studies of quantum critical behavior in spin-dimer systems.

\acknowledgments
We are grateful to ESRF and ILL for granting the beam time, and acknowledge experimental assistance of Andy Fitch, Carolina Curfs, Adrian Hill, Dmitry Batuk, and Oleg Janson at the ID31 beamline of ESRF. We also acknowledge Horst Borrmann and Yurii Prots for laboratory XRD measurements, Stefan Hoffmann for his help with thermal analysis, Oleg Janson for fruitful discussions, and Martin Rotter for the careful reading of the manuscript. A.T. was supported by Alexander von Humboldt Foundation.

%\bibliography{CuCl-Ta}

\begin{thebibliography}{63}
\bibitem [{f1()}]{f1}%
  \BibitemOpen
  \href@noop {} {}\bibinfo {note} {Weak unindexed reflections at
  $2\theta=22$~deg and 31.5~deg are likely experimental artifacts arising,
  e.g., from the cryostat. We did not observe these extra reflections in the
  D2B and synchrotron data.}\BibitemShut {Stop}%
\end{thebibliography}

\newpage
\renewcommand{\thefigure}{S\arabic{figure}}
\setcounter{figure}{0}
\setcounter{table}{1}
\renewcommand{\thetable}{S\arabic{table}}

\begin{widetext}
\begin{center}

{\large
Supplementary information for 
\smallskip

\textbf{(CuCl)LaTa$_2$O$_7$ and quantum phase transition\\ in the (CuX)LaM$_2$O$_7$ family (X = Cl, Br; M = Nb, Ta)}}
\medskip

A. A. Tsirlin, A. M. Abakumov, C. Ritter, and H. Rosner
\end{center}
\medskip

\begin{figure}[!h]
\includegraphics[width=11cm]{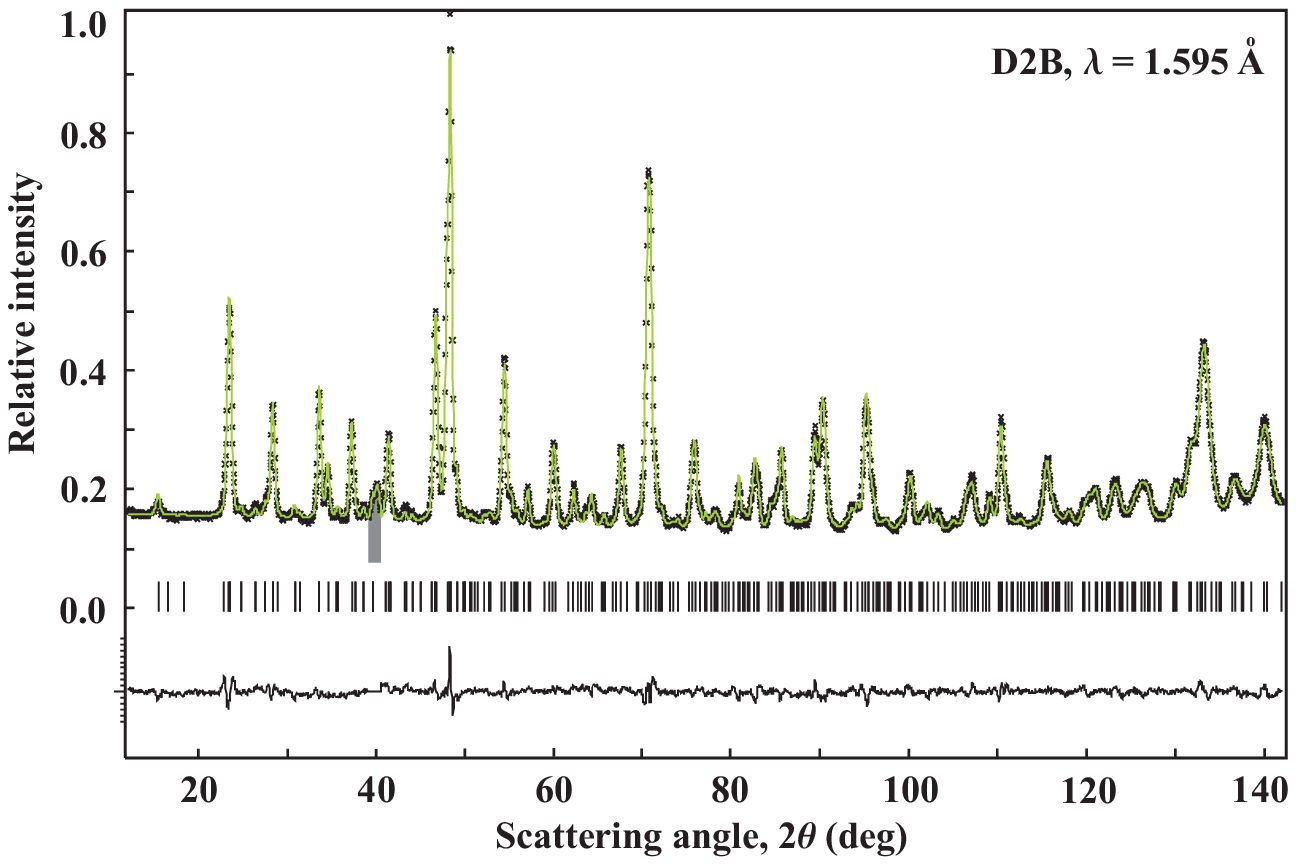}
\begin{minipage}{14cm}
\caption{\label{fig:s1}\normalsize
Rietveld refinement of the D2B neutron data collected at 1.8~K. The excluded region around $2\theta=40$~deg is due to the cryostat window. Refinement residuals are $R_I=0.021$, $R_p=0.024$, and $R_{wp}=0.030$.
}
\end{minipage}
\end{figure}
\bigskip

\begin{figure}[!h]
\includegraphics[width=11cm]{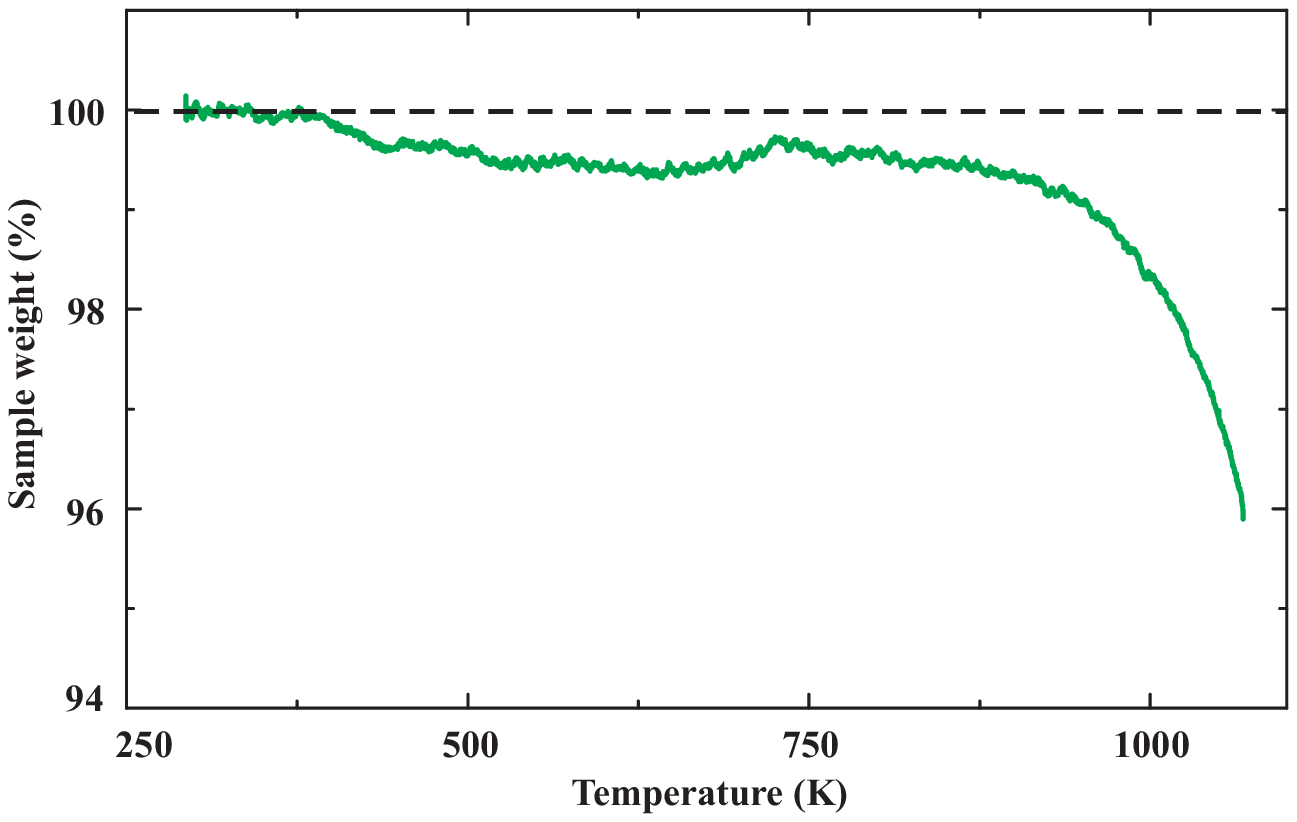}
\begin{minipage}{14cm}
\caption{\label{fig:s2}\normalsize
Thermogravimetric data for (CuCl)LaTa$_2$O$_7$. Note the onset of the weight loss around 850~K.
}
\end{minipage}
\end{figure}

\end{widetext}

\end{document}